\documentclass{nic-series}
\begin{document} 
\title{Reduction to the simplest -- The complexity of minimalistic heteropolymer models}
\author{Michael Bachmann \and Wolfhard Janke}
\institute{Institut f\"ur Theoretische Physik and Centre for Theoretical Sciences (NTZ), \\
        Universit\"at Leipzig, 04109 Leipzig, Germany\\
        \email{\{bachmann, janke\}@itp.uni-leipzig.de}}
\maketitle
\begin{abstracts}
Simple coarse-grained hydrophobic-polar models for heteropolymers as the lattice HP and the
off-lattice AB model allow a general classification of characteristic behaviors
for hydrophobic-core based tertiary folding.
The strongly reduced computational efforts enable one
to reveal systematically the thermodynamic properties of comparatively long sequences in a wide temperature
range of conformational activity. 
Based on a suitable cooperativity parameter, characteristic folding channels and free-energy 
landscapes, which have strong similarities with realistic folding paths, can be analysed. 
\end{abstracts}
\section{Introduction}
The understanding of the sequence-dependent stability of native protein structures, the interplay of
sequence and conformation, and the classification of conformational phases are essential aspects in 
protein folding studies. By means of computer simulations of detailed microscopic models with 
modern algorithms, great progress has been achieved in the past few years in predicting native 
structures for exemplified amino acid sequences. Also, the relation between conformational cooperativity
and thermodynamic activity in the folding process has been subject of numerous selective
studies. Nonetheless, a general framework, which would allow the classification of characteristic 
global folding events on the basis of an appropriate ``order'' or cooperativity parameter, is still lacking. 
Another difficult problem is the systematic analysis of the influence of the amino acid order in the
sequence on folding kinetics and native fold. 
Since these questions cannot or only partially be answered with all-atom protein models, it
is necessary to increase the level of abstraction in appropriate modelling. It is known that
most of these problems are related to the so-called tertiary folding process which concerns 
almost all amino acid residues in a cooperative manner, in contrast to the much more localised 
secondary structure segments. The main idea is that
tertiary folding is strongly related to the hydrophobic effect. Many proteins fold spontaneously
in an aqueous solution and the hydrophobicity of the distinguished side chains is mainly responsible 
for the position of the side chains in the native fold. Polar residues favor contact with the 
polar environment and screen more hydrophobic monomers, which typically form a compact core
in the interior of the protein structure. 
In connection with volume exclusion, it is the hydrophobic
force that enters into minimalistic hydrophobic-polar models.
\section{The Hydrophobic-Polar (HP) Lattice Model}
The simplest hydrophobic-core based approach is the HP lattice model~\cite{bjdill1}, where 
in the original form only nearest-neighbor hydrophobic monomers, being nonadjacent along the
self-avoiding chain, attract each other. 
A remarkable result obtained from exactly enumerating the entire
sequence and conformation space of HP lattice proteins in this parameterization~\cite{bjtang1,bjtang2,bjsbj1} 
is that only a small fraction of
possible sequences 
is nondegenerate, i.e., for these only one conformation representing the 
``native'' state exists. This is interpreted as indication of stability in the funnel-like free-energy
landscape: There is an energy gap towards higher-energetic states. Such sequences are called 
{\em designing}. On the other hand, it is interesting that the number of designable 
conformations, i.e., structures being native folds of designing sequences, is also very small,
typically even smaller than the number of designing sequences.~\cite{bjsbj1}
This means that several designing sequences fold into the same native conformation.   
In studies of thermodynamic activity accompanying the folding process, it turned out that,
depending on the sequence, the folding trajectory can pass highly compact intermediate 
globular states. 
Favoring hydrophobic contacts energetically, however, hydrophobic-core folds are
not necessarily maximally compact.~\cite{bjsbj1,bjbj1} Identifying low-lying energetic states
and analyzing thermodynamic properties of HP lattice proteins is not straightforward and requires 
sophisticated computational methods.~\cite{bjbj1,bjhsu1} Although the tendency of the HP model
to favor folding paths with intermediary conformational phases is seen as sign of too little
cooperativity, its power lies in the fact that sequences with far more than 100 monomers can be
analyzed. These sequences are much longer than what can be studied by means of other models which are
not knowledge-based, i.e., not biased towards a certain target structure. 
\section{Getting Rid of the Lattice: A Minimal Continuum Model}
\begin{figure}
\centerline{\epsfxsize=8.0cm \epsfbox{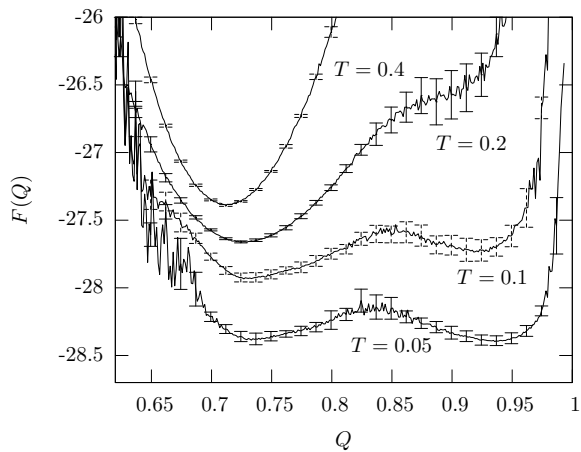}}
\caption{\label{fig:bjtwo} Free-energy landscapes $F(Q)$ at different temperatures 
for the sequence $BA_6BA_4BA_2BA_2B_2$, exhibiting typical two-state folding behavior.~\cite{bjsbj2}}
\end{figure}

A straightforward off-lattice generalization of the HP model is the AB model ($A$: hydrophobic, 
$B$: polar), which was originally 
proposed for heteropolymers in two dimensions.~\cite{bjstill1}
Volume exclusion and effective inter-residue interactions are modeled by Lennard-Jones potentials,
and a ``ferromagnetic'' bending energy models effectively stiffness and torsional barriers
in the three-dimensional representation.~\cite{bjbaj1,bjsbj2} For studies of folding channels,
we define an angular overlap parameter of torsional and bond angles. Comparing
a conformation ${\bf X}$ with the native conformation ${\bf X_0}$, this similarity parameter is
defined by $Q({\bf X})=1-d({\bf X})$, where $d({\bf X})$ is a measure or the angular deviation of 
${\bf X}$ from ${\bf X_0}$.~\cite{bjbaj1,bjsbj2} Restricting the canonical partition function 
at temperature $T$ to the ensemble of conformations with overlap $Q$, 
$Z(Q)= \int {\cal D}{\bf X}\, \delta(Q-Q({\bf X}))\,\exp\{-E({\bf X})/k_BT\}$, the overlap free energy 
is given by $F(Q)=-k_BT\ln Z(Q)$. 
Figure~\ref{fig:bjtwo} shows the overlap free-energy landscapes for the 20-mer $BA_6BA_4BA_2BA_2B_2$,
which behaves like a typical two-state folder. For permuted sequences, we also find characteristic
folding through intermediate conformations and metastability.~\cite{bjsbj2} 
\section{Concluding Remarks}
Protein folding is a highly complex process where microscopic details are of essential importance.
Nonetheless, the effective hydrophobic force which is responsible for the tertiary fold is of rather
mesoscopic nature, since it is due to the interaction with the bath of surrounding solvent molecules.
For this reason, appropriate effective coarse-grained models should allow a qualitative description 
of characteristic features accompanying conformational folding transitions. By means of minimalistic
standard hydrophobic-polar models we have shown that even at this level of abstraction typical properties of native
folds and cooperative effects in the folding process can be described qualitatively.   
\section*{Acknowledgments}
We thank H.\ Ark{\i}n, R.\ Schiemann, S.\ Schnabel, T.\ Vogel, A.\ Kallias, J.\ Schluttig, 
and C.\ Junghans for cooperation.
This work is partially supported by the DFG (German Science Foundation) under Grant
No.\ JA 483/24-1 and the computer time Grant No.\ hlz11 of the John von Neumann Institute for 
Computing (NIC), Forschungszentrum J\"ulich.

\end{document}